\documentclass[twocolumn,showpacs]{revtex4}

\def\nh4{$\alpha$-(BEDT-TTF)$_2$NH$_4$Hg(SCN)$_4$}

\usepackage{graphicx}
\usepackage{dcolumn}
\usepackage{amsmath}
\begin{document}

%\twocolumn[\hsize\textwidth\columnwidth\hsize\csname
%@twocolumnfalse\endcsname
\title{Absence of field-induced charge-density wave states in 
(Per)$_2$Pt(mnt)$_2$}
\author{R.~D.~McDonald, N.~Harrison, P.~Goddard,
J.~Singleton and X.~Chi
}
\affiliation{National High Magnetic Field Laboratory, LANL, MS-E536, Los
Alamos, New Mexico 87545}
%\date{\today}
%\maketitle

%]\narrowtext
\maketitle 

Graf {\it et al.}~\cite{graf1} recently attributed features
in the magnetic-field-dependent longitudinal resistance of
(Per)$_2$Pt(mnt)$_2$ to a cascade of
field-induced charge-density waves
(FICDWs).  Here we show that a quantitative
magnetotransport analysis reveals orbital quantization to be absent,
disproving the presence of FICDWs.  The conduction is instead
dominated by the sliding CDW collective mode at low temperatures.

It has been suggested~\cite{lebed1} that
FICDWs occur in charge-density wave (CDW) systems in
strong magnetic fields when orbital quantization 
facilitates nesting
of quasi-one-dimensional Zeeman-split bands.  
The free energy is
minimised at low integral Landau subband 
(LS) filling factors $\nu$ by
the formation of a Landau gap at the 
Fermi energy~\cite{lebed1}. 
Hence, as is the case in field-induced spin-density 
wave states (FISDW)~\cite{chaikin1}, orbital quantization 
is implicit in FICDW
formation, yielding a Hall conductivity
$\sigma_{xy}\approx 2\nu e^2/ah$ (where $a\sim$~20~\AA~ is the layer
spacing) and a longitudinal conductivity
$\sigma_{xx}\approx\sigma_0\exp[-\Delta/k_{\bf B}T]$ that is very
small and thermally activated.
Since $\sigma_{xy}/\sigma_{xx}\equiv\rho_{xy}/\rho_{xx}\gg$~1, conditions
are ripe for the quantum Hall effect, where $\rho_{xy}\approx
1/\sigma_{xy}\approx ah/2\nu e^2$, while the longitudinal resistivity,
\begin{equation}\label{FICDW}
    \rho_{xx}\approx\sigma_{xx}/\sigma_{xy}^2\approx(ah/2\nu
    e^2)^2\sigma_0\exp[-\Delta/k_{\bf B}T],
\end{equation}
is comparitively small and decreases with decreasing 
temperature $T$.  The
data of Ref.~\cite{graf1} are inconsistent with
this scenario: the apparent $\rho_{xx}$ has the inverse
$T$-dependence to Eq.~\ref{FICDW} and a magnitude
$\rho_{xx}\sim$~0.5~$\Omega$m at 0.5~K (estimated using the given crystal
dimensions~\cite{graf1}) that is much too large.  
Since the Hall
resistivity cannot exceed
$\rho_{xy}\sim$~30~$\times$~10$^{-6}$~$\Omega$m, this implies
$\rho_{xy}/\rho_{xx}\ll$~10$^{-4}$, incompatible with LS
formation.

\begin{figure}[htbp]
\centering
\vspace{-0.15in}
\includegraphics[width=7.5cm]{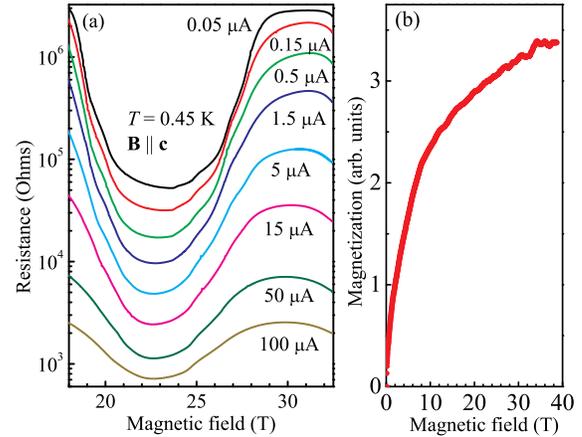}
\caption{(a)~Resistance versus field for 
(Per)$_2$Pt(mnt)$_2$ at various currents.
(b)~Magnetization $M$ of many
randomly oriented
(Per)$_2$Pt(mnt)$_2$ crystals at $T=0.50$~K;
$M$ does not saturate by $B \approx 40$~T.}
\label{fig1}
\end{figure}

By neglecting to change their sample current
from $I=1~\mu$A, Graf {\it et
al.} also failed to observe the highly non-linear 
behavior of the
resistivity. Fig.~\ref{fig1}(a) shows 
our measurements of the
resistance of a similar single 
crystal of (Per)$_2$Pt(mnt)$_2$
versus field for several currents.
These data show that for $I \sim 1~\mu$A
and $T \approx 0.5$~K (conditions similar to Ref.~\cite{graf1}),
the conductivity is clearly in
the non-linear regime, where it is 
dominated by the sliding CDW
collective mode, {\it at all magnetic fields} $\leq 33$~T.  
What Graf {\it et al} term 
``magnetoresistance''
is therefore not strictly magnetoresistance at all, 
but mostly the
consequence of field-induced changes in the 
threshold electric field
$E_{\rm t}$ required to 
depin the CDW from the lattice.  
Thus, the combined
data of Graf {\it et al.}~\cite{graf1} and
Fig.~\ref{fig1}(a) correspond to a scenario in
which $\sigma_{xy}\approx$~0, giving a total resistivity
\begin{equation}\label{CDW}
\rho_{xx}=(\sigma_0\exp[-\Delta/k_{\rm B}T]+j/E_{\rm t})^{-1},
\end{equation}
in which the thermally-activated conductivity and the CDW collective
mode occur in parallel.  Here, $j$ is the current density; note that
$E_{\rm t}$ may itself depend on $T$.
Such behavior is totally inconsistent with the
predictions of Eq.~\ref{FICDW};
rather than being caused by FICDW phases~\cite{lebed1}, 
the steps in
$\rho_{xx}$ probably correspond to field-induced changes in
$E_{\rm t}$.

The cooperative dimerization of the Pt spins in (Per)$_2$Pt(mnt)$_2$
can easily provide a mechanism for additional phase transitions or
changes in $E_{\rm t}$ compared to
(Per)$_2$Au(mnt)$_{2}$~\cite{mcdonald1}.  
The Pt spins couple strongly
to both the CDW, via distortions of the crystal lattice, and the
magnetic field, as shown by the 
fact that they dominate the total
magnetic susceptibility (i.e. the 
change in free energy of the
composite system as a function of field) (Fig.~\ref{fig1}(b)).  
Their effect on the
phase diagram is likely to be significant until all spins are fully
aligned by a field $B\gtrsim$~40~T (Fig.~\ref{fig1}(b)).

%% 
%This work is supported by US Department of Energy (DOE) grant \#LDRD-DR
%20030084 and was performed under the auspices of the National Science
%Foundation, the DOE and the State of Florida. 

\end{document}